\documentclass[aps,prd,twocolumn,showpacs,amsmath,amssymb,nofootinbib,eqsecnum, preprintnumbers]{revtex4-2}
\usepackage{graphicx}
\usepackage{epstopdf}
\usepackage{epsfig}
\usepackage{xcolor}
\usepackage{verbatim}

\def\appendix{{\newpage\section*{Appendix}}\let\appendix\section%
        {\setcounter{section}{0}
        \gdef\thesection{\Alph{section}}}\section}

\newcommand{\be}{\begin{equation}}
\newcommand{\ee}{\end{equation}}
\newcommand{\bear}{\begin{eqnarray}}
\newcommand{\eear}{\end{eqnarray}}
\newcommand{\ba}{\begin{array}}
\newcommand{\ea}{\end{array}}

\newcommand{\tcb}{\textcolor{blue}}
\newcommand{\tcr}{\textcolor{red}}
\newcommand{\tco}{\textcolor{orange}}

\begin{document}

\title{Thermodynamic relation of accelerating black holes in anti-de Sitter spacetime}
\author{Chong Oh Lee}
\email{cohlee@gmail.com}
\affiliation{Center of Liberal Arts, Wonkwang University, Iksan, Jeonbuk 54538, Republic of Korea}

\begin{abstract}
We consider accelerating black holes in four-dimensional anti-de Sitter spacetime
and investigate the extremality relations by examining perturbative corrections to both the entropy of black holes and their extremality bounds.
It is shown that the so-called Goon and Penco (univeral) relation is also valid for accelerating black holes.
Furthermore, it is found that an appropriate matching condition is necessary  between the perturbation parameter of the
relation with the perturbative correction to such black holes, and the parameter with information
about the conical deficits on the north and south poles with the cosmic string tensions in order to
satisfy the extremality relations.
This result indicates that the extremality (univeral) relation of a class of accelerating black holes holds
exhibit behavior similar to the Weak Gravity Conjecture.
\end{abstract}
\maketitle
\section{Introduction}
It was proposed that the Weak Gravity Conjecture (WGC) aims to provide insights into the nature of gravity
and gauge theories in the context of string theory and quantum gravity~\cite{Vafa:2005ui}.
The WGC suggests that in any consistent theory of quantum gravity,
the gravitational force should be weaker than the force of some lighter, charged particle
or field~\cite{Arkani-Hamed:2006emk}. In particular, it revelas that there should exist a particle with a charge ${\cal Q}$ and a mass ${\cal M}$
such that the force it experiences due to gravity is weaker than the force it experiences due to its gauge interaction i.e., ${\cal Q}<{\cal M}$.
However, when the extremal black hole (BH) saturates the bound for a charged BH,
the relation of the charge ${\cal Q}$ and the mass ${\cal M}$ is written as
\be\label{QM}
{\cal M}/{\cal Q}\leq 1.
\ee
This conjecture is especially important in the context where global symmetries are absent~\cite{Banks:2010zn}
and is based on the principles of quantum gravity~\cite{Harlow:2022ich}. Furthermore, the naked singularity would  violate the inequality~(\ref{QM}),
aligning with the WGC, but it is prohibited by cosmic censorship.
A possible solution to this problem is to incorporate correction terms into the action~\cite{Cheung:2018cwt},
which have been extensively studied in the context of higher-derivative correction terms
~\cite{Cheung:2018cwt,Kats:2006xp,Reall:2019sah,Cheung:2019cwi,Ma:2023qqj}.
Garrett Goon and Riccardo Penco suggested that the universal relation between corrections to entropy, extremality, and their extremality bounds
in the context of BH physics typically refers to how small perturbations or corrections
to the properties of a BH affect its entropy and the conditions under which it remains extremal
i.e., at the threshold where a BH becomes a naked singularity if exceeded~\cite{Goon:2019faz}.
This relation was generalized by examining its applications to a variety of physical systems
~\cite{Wei:2020bgk,Chen:2020bvv,Chen:2020rov,Ma:2020xwi,Sadeghi:2020ciy,Sadeghi:2020xtc,McInnes:2021frb,
McInnes:2021zlt,Noumi:2022ybv,Etheredge:2022rfl,Ko:2023nim,Ko:2024ymy}, and
other related works can be found in
~\cite{Cano:2019ycn,Cremonini:2019wdk,Arkani-Hamed:2021ajd,Sadeghi:2020ntn,McPeak:2021tvu,Sadeghi:2022xcr,Xiao:2023two}.

The concept of accelerating BHs was first introduced by William Kinnersley and Milton Walker,
where they presented the so-called $C$-metric~\cite{Kinnersley:1970zw,Plebanski:1976gy,Griffiths:2005qp}.
This metric can be written as the solution to Einstein's field equations with a negative cosmological constant~\cite{Plebanski:1976gy},
and describes a pair of BHs accelerating away from each other under the influence of some force.
The $C$-metric is foundational for understanding the acceleration of BHs and includes conical singularities
that represent the force causing the acceleration.
Detailed calculations of the stress tensor were provided, and its physical interpretation
in the boundary theory was discussed by analyzing the holographic stress tensor for accelerating BHs
and examining how the acceleration modifies the usual AdS/CFT dictionary ~\cite{Anabalon:2018ydc}.
In the context of charged accelerating AdS BHs, the works explored
the interaction between modified gravity theories and BH physics within the framework of $f(R)$ gravity~\cite{Belhaj:2020lai},
and analyzed the factors affecting the efficiency of these heat engines~\cite{Zhang:2018hms}.
Their generalized version for the thermodynamic properties and holographic dualities in lower-dimensional AdS spaces
has been studied in~\cite{Zhang:2018hms,Arenas-Henriquez:2023hur,Cisterna:2023qhh, Tian:2023ine}.

The paper is organized as follows: In the next section, we will briefly review the universal relation between corrections to entropy
and extremality and their extremality bounds, as well as accelerating BH
with a negative cosmological constant in four-dimensional spacetime.
Next, we will apply perturbative corrections to the entropy and extremality bounds of accelerating BHs,
demonstrating that the universal relation is confirmed within the holographic framework of these BHs.
Finally, we summarize our results and give our discussion in the last section.

\section{Extremality relation}
For both the entropy of BHs and their extremality bounds,
universal relation from considering perturbative correction to the theory was suggested in~\cite{Goon:2019faz}
\begin{equation}\label{ER}
\mkern-9mu\frac{\partial M_{\rm ext}(\vec{Q},\epsilon)}{\partial \epsilon}
=\lim_{M\to M_{\rm ext}(\vec{Q},\epsilon)}
\mkern-9mu-T\left (\frac{\partial S(M,\vec{Q},\epsilon)}{\partial \epsilon}\right )_{M,\vec{Q}},
\end{equation}
where $\vec{Q}$ and $\epsilon$ are additional quantities of the BH thermodynamics such as charge, angular momentum or other quantities,
and perturbation parameter. Here $M_{\rm ext}$, $M$, $T$, and $S$ denote the extremality mass, mass, temperature
and entropy of the BH respectively after the correction.
From extremality relation (\ref{ER}), after taking approximate relations with holding at leading order in a perturbative expansion,
one can lead to a leading-order expression
\begin{equation}\label{LO}
\Delta M_{\rm ext}(\vec{Q})
\approx-T_{0}(M,\vec{Q})\Delta S(M,\vec{Q})
\mid_{M\approx M^{0}_{\rm ext}(\vec{Q})},
\end{equation}
where $\Delta S(M,\vec{Q})$ and $\Delta M_{\rm ext}(\vec{Q})$ represent
the leading corrections to the entropy of a state with fixed $M$, $\vec{Q}$, and the extremality bound, respectively.

The metric with accelerating BH in the four-dimensional anti-de Sitter (AdS)
is able to be described by \cite{Griffiths:2005qp,Podolsky:2002nk,Hong:2003gx}
\begin{equation}\label{AM}
ds^2=\frac{1}{p}\left[-fdt^2+\frac{dr^2}{f}+r^2\left(\frac{d\theta^2}{q}+q\sin^2 \theta\frac{d\phi^2}{K^2}\right)\right],
\end{equation}
with
\begin{equation}
\begin{aligned}
p(r,\theta)&=1+Ar\cos\theta, \quad  q(\theta)= 1+2mA\cos\theta,\\
f(r)&=(1-A^2r^2)\bigg(1-\frac{2m}{r}\bigg)+\frac{r^2}{l^2},
\end{aligned}
\end{equation}
where the metric function $f(r)$ is the BH of the solution with the mass parameter $m$ (the mass of the BH $M=m\sqrt{1-A^2l^2}/K$)~~\cite{Anabalon:2018ydc},
$A$ is an acceleration parameter, and $l$ denotes the four-dimensional cosmological parameter (the cosmological constant $\Lambda=-3/l^2$).
The parameter $K$ due to requiring $2mA<1$ to preserve the metric signature is naturally introduced
so that it encodes information about the conical deficits on the north and south poles
with the cosmic string of tensions ($\mu_{+}$ at the north pole and $\mu_{-}$ at the south pole)~\cite{Appels:2017xoe,Anabalon:2018ydc}
\be\label{ST}
\mu_{\pm}=\frac{\delta_{\pm}}{8\pi}=\frac{1}{4}\left(1-\frac{q(\theta_\pm)}{K}\right)
=\frac{1}{4}\left(1-\frac{1\pm 2mA}{K}\right),
\ee
where $\delta_{+}$ is the conical deficit at the north pole and $\delta_{-}$ is the conical deficit at the south pole.
In particular, when the BH with conical deficits has a regular pole at the north axis, its conical deficit $\delta_{+}$
becomes zero and the parameter $K$ is determined as $K=1+2mA$. When such BH has the conical deficit at the south pole,
from Eq. (\ref{ST}) its conical deficit $\delta_{-}$ is equal to $8\pi m A/K$
and the cosmic string of tensions on south axis $\mu_{-}$ is determined as $mA/K$.
Here, the tension of the string exerts a force to accelerate the BH,
similar to Newton's second Law (the tension of the string behaves
as if the acceleration of the object is directly proportional to the net force acting on the object).

In fact, one can interpret the tension of string
through analysing the equations of motion for a BH with an cosmic string vortex.
For example, a Schwarzschild BH with conical deficits is given as~ \cite{Appels:2017xoe}
\be\label{SB}
ds^2=-Fdt^2+\frac{dr^2}{F}+r^2\left(d\theta^2+\sin^2\theta\frac{d\phi^2}{K}\right),
\ee
with $F=1-2m/r$.
When $K=1$, the tension of string becomes zero and the BH has a regular pole.
Furthermore, the tension along either polar axis becomes equal
and both the two poles are simultaneously regularized. On the other hand, when $K\neq1$,
there is a conical defect along the axis of revolution in the BH and the tension of string becomes non-zero.
Thus, the parameter $K$ has relation with the tension of string $\mu$ and the conical deficit $\delta$ which given as
\be\label{TS}
\mu=\frac{\delta}{8\pi}=\frac{1}{4}\left[1-\frac{1}{K}\right].
\ee

We consider the cosmological constant with the perturbation parameter $\epsilon$ and from the metric Eq. (\ref{AM}) obtain the metric function
\be\label{PM}
f(r)=(1-A^2r^2)\bigg(1-\frac{2m}{r}\bigg)+\frac{(1+\epsilon)r^2}{l^2},
\ee
which becomes the shifted mass parameter $m$
\be\label{SM}
m=\frac{r}{2}\left(1-\frac{r^2(1+\epsilon)}{L^2(A^2r^2-1)}\right),
\ee
by using $f(r)=0$.

One may take the entropy to be one quarter of the horizon area $\cal A$
\be
S=\frac{\cal A}{4}=\frac{\pi r_+^2}{K(1-A^2r_+^2)},
\ee
which satisfies $f(r_+)=0$ and leads to
\be\label{RP}
r_+=\frac{\sqrt{K}\sqrt{S}}{\sqrt{\pi+A^2KS}}
\ee

From the metric Eq. (\ref{PM}) one may compute the temperature~\cite{Anabalon:2018ydc}
\be\label{MT}
T=\frac{f'(r_+)}{4\pi \sqrt{1-A^2 l^2}}=\frac{(1-A^2r^2)^2+r_+^2(3-A^2r_+^2)(1+\epsilon)/l^2}{4\pi\sqrt{1-A^2 l^2}r_+(1-A^2r_+^2)},
\ee
and substituting Eq. (\ref{RP}), one finds
\be\label{PT}
T=\frac{\pi^2+ KS(3\pi+2A^2KS)(1+\epsilon)/l^2}{4\pi^2\sqrt{1-A^2 l^2}\sqrt{K}\sqrt{S}\sqrt{\pi+A^2KS}}.
\ee

Note that when $m=0$, the BH horizon vanishes, generally leaving pure AdS spacetime in Rindler-type coordinates.
However, in the case of accelerating BHs, the time coordinate is not AdS time but is instead rescaled by a factor of $\sqrt{1-A^2 l^2}$.
Furthermore, one may normalize one's time coordinate to correspond to the $time$ experienced by an asymptotic observer.
Although this can be a somewhat ambiguous concept in AdS, when combined with spherical asymptotic spatial coordinates,
this scaling implies that the correct time coordinate is not $t$, but rather $\tau=t\sqrt{1-A^2 l^2}$.
This results in a rescaling of the time coordinate in Eq. (\ref{AM}) and
the temperature (\ref{MT}) contains the rescaling factor of $\sqrt{1-A^2 l^2}$~\cite{Anabalon:2018ydc}
unlike the usual form of Hawking temperature $T=f'(r_+)/(4\pi)$.

After substituting Eq. (\ref{RP}) into Eq. (\ref{SM}), the shifted mass $m$ is given as
\be\label{RdM}
m=\frac{\sqrt{K}\sqrt{S}(\pi+KS(1+\epsilon)/l^2)}{2\pi\sqrt{\pi+A^2KS}},
\ee
which leads to
\be\label{EPS}
\epsilon=\frac{2\pi\sqrt{K}\sqrt{S}\sqrt{\pi+A^2KS}l^2M}{K^2S^2}-\frac{\pi l^2}{KS}-1,
\ee
and taking partial derivative of Eq. (\ref{EPS}) with fixed mass parameter $m$, one gets
\be\label{PEPS}
\left(\frac{\partial\epsilon}{\partial S}\right)_{m}=-\frac{\pi^2l^2+KS(3\pi+2A^2KS)(1+\epsilon)}{2KS^2(\pi+A^2KS)}
\ee

From Eq.(\ref{PT}) and Eq.(\ref{PEPS}), one may compute
\be\label{LEFT}
-T\frac{\partial S}{\partial\epsilon}=\frac{S^{3/2}\sqrt{K(\pi+A^2KS)}}{2\pi^2\sqrt{1-A^2l^2}l^2}.
\ee

One takes partial derivative of Eq. (\ref{RdM}) with fixed temperature $T$
\be\label{RIGHT}
\left(\frac{\partial m}{\partial\epsilon}\right)_{T}=\frac{(KS)^{3/2}}{2\pi\sqrt{\pi+A^2KS}l^2}.
\ee

\begin{table*}[hbt!]
\caption{The parameter $K(\epsilon)$ for the various variable $A^2l^2$.}
\begin{tabular}{ |c|c|c|c|}
\hline
$A^2L^2$          &$\epsilon=10^{-10}$          &$\epsilon=0.5$          &$\epsilon=1$       \\
\hline
\tcr{0}           &\tcr{1.00000000}             &\tcr{1.00000000}        &\tcr{1.00000000}   \\
0.01              &1.00001667                   &1.00003000              &1.00003334         \\
0.1               &1.00168000                   &1.00280000              &1.00336000         \\
\tco{0.7}         &\tco{1.13909000}             &\tco{1.23476000}        &\tco{1.27891000}   \\
\tcb{0.9}         &\tcb{1.48400000}             &\tcb{1.81429000}        &\tcb{1.95003000}   \\
$1-10^{-10}$      &35355.3391                   &51527.8000              &57206.1403         \\
\hline
\end{tabular}
\end{table*}

After taking $T=0$ from Eq.(\ref{PT}), one can obtain the entropy $S_{\rm E}$ at the extremal condition
\be\label{EE}
S_{\rm E}=\frac{\pi(-3\sqrt{1+\epsilon}+\sqrt{9-8A^2l^2+9\epsilon})}{4A^2K\sqrt{1+\epsilon}}
\ee

By substituting Eq.(\ref{EE}) into Eq. (\ref{LEFT}) and Eq. (\ref{RIGHT}), one can get
\be\label{ETS}
\lim_{m\to m_{\rm E}}\left(-T\frac{\partial S}{\partial\epsilon}\right)_{m}=\frac{(-3+\sqrt{\eta})^{3/2}\sqrt{1+\sqrt{\eta}}}{32\sqrt{1-A^2 l^2}A^3Kl^2},
\ee
and
\be\label{EM}
\left(\frac{\partial m_{\rm E}}{\partial\epsilon}\right)_{T}=\frac{(-3+\sqrt{\eta})^{3/2}\sqrt{1+\sqrt{\eta}}}{8(1+\sqrt{\eta})^{3/2}\sqrt{1+\epsilon}A^3l^2}
\ee
where $\eta=(9-8A^2l^2+9\epsilon)/(1+\epsilon)$. Here, Eq. (\ref{ETS}) is depended on the parameter $K$ and the factor of $\sqrt{1-A^2l^2}$
while Eq. (\ref{EM}) is independent of them. This requires additional condition which satisfies
the universal relation from considering perturbative correction to accelerating BH, Eq. (\ref{ETS}).
From Eq. (\ref{EM}) into Eq. (\ref{ER}), one can get
\be
K\sqrt{1-A^2l^2}=\frac{1}{4}\left(1+\frac{\sqrt{9-8A^2l^2+9\epsilon}}{\sqrt{1+\epsilon}}\right),
\ee
through the matching  between the left-hand side of Eq. (\ref{ER}) and the right-hand side of Eq. (\ref{ER})
and finally leads to following the relation
\be\label{K}
K(\epsilon,~A^2l^2)=\frac{1}{4\sqrt{1-A^2l^2}}\left(1+\frac{\sqrt{9-8A^2l^2+9\epsilon}}{\sqrt{1+\epsilon}}\right).
\ee
The parameter $K$($\epsilon$, $A^2l^2$), as the function of the perturbation parameter $\epsilon$
and the variable $A^2l^2$, seems to arise naturally since $K$ is linked to the string tension,
that behaves as if the acceleration of the object is directly proportional to the net force acting on the object, as previously mentioned.

As you see in FIG. 1, $K\rightarrow1$ as $A^2 l^2$ goes to 0 for Eq. (\ref{K}) with a given perturbation parameter $\epsilon$.
Furthermore, when $A^2l^2=0$, accelerating BH in the four-dimensional AdS is recovered
to the four-dimensional AdS BH with conical deficits~\cite{Appels:2017xoe}:
\be\label{CBH}
ds^2=-Gdt^2+\frac{dr^2}{G}+r^2(d\theta^2+\sin^2\theta)\frac{d\phi^2}{K^2},
\ee
where
\be
G=\left(1-\frac{2m}{r}\right)+\frac{r^2}{l^2},
\ee
via the metric Eq. (\ref{AM}). Then from the Eq. (\ref{K}) when $K$ is equal to 1, the four-dimensional AdS BH with conical deficits
reduces the four-dimensional AdS BH. This result implies that by imposing the universal relation
on the four-dimensional AdS BH with conical deficits, the string tension becomes zero and the BH should have a regular pole.
These results are consistent with that of \cite{Wei:2020bgk,Chen:2020rov,Sadeghi:2020ciy,Sadeghi:2020xtc}.
Furthermore, $K$ becomes infinity as $A^2 l^2$ approaches 1 as shown in FIG. 1.
In this limit, the acceleration horizons disappear since $M~(M=m\sqrt{1-A^2l^2}/K$) vanishes.
As we mentioned, $K$ is closely related to the string tension, and this tension becomes infinite as $K$ approaches infinity.
This implies the BH that experiences this tension undergoes infinite acceleration and moves at infinite velocity, which is unphysical.
Thus, the universal relation seems not to work in the limit of the vanishing of acceleration horizons.

\begin{figure}[!htbp]
\begin{center}
{\includegraphics[width=7cm]{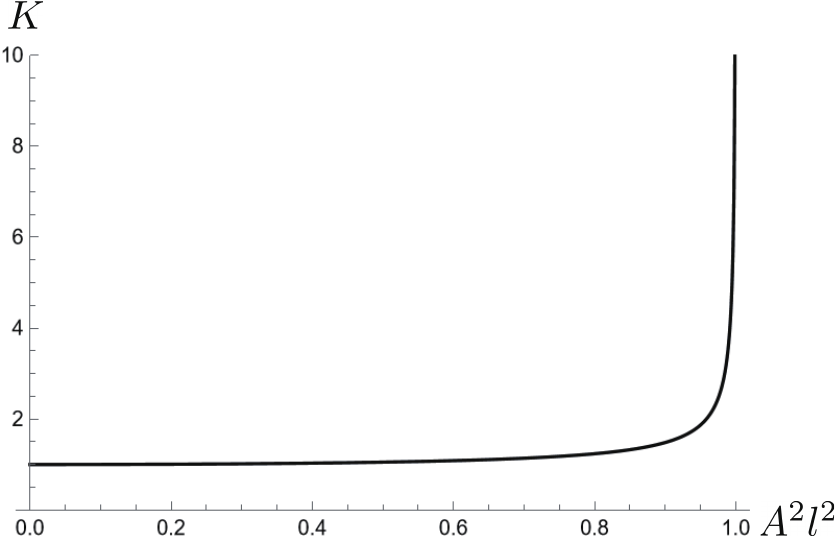}}
\vspace{-0.6cm}
\end{center}
\caption{{\footnotesize From Eq. (\ref{K}), plot of the parameter $K$($A^2l^2$) as the function of $A^2l^2$ for $\epsilon=10^{-5}$.}}
\label{figI}
\end{figure}

Shapes of the parameter $K(\epsilon)$ for various values of the variable $A^2l^2$ are depicted in FIG. 2.
The parameter $K(\epsilon)$ monotonically as the perturbation parameter $\epsilon$ increases.
As the variable $A^2l^2$ grows, the value of the parameter $K(\epsilon)$ also increases as shown in FIG. 2, and in TABLE I.

\begin{figure}[!htbp]
\begin{center}
{\includegraphics[width=7cm]{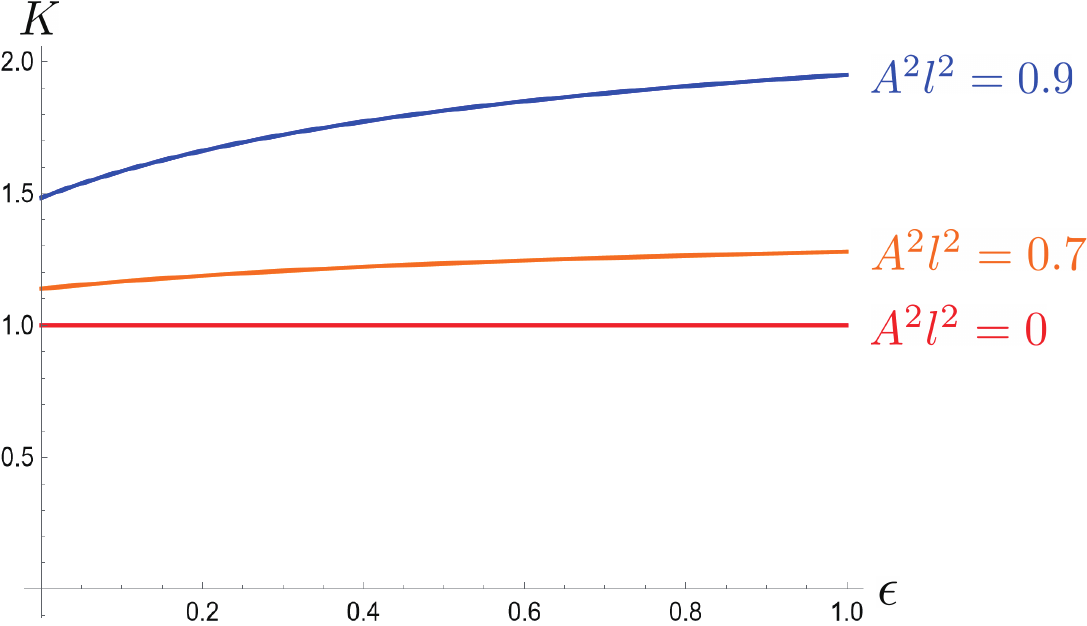}}
\vspace{-0.6cm}
\end{center}
\caption{{\footnotesize From Eq. (\ref{K}), plot of the parameter $K(\epsilon)$as the function of $\epsilon$
for $A^2l^2=0$ (solid curve red), $A^2l^2=0.7$ (solid curve orange), and $A^2l^2=0.9$ (solid curve blue), respectively.}}
\label{figI}
\end{figure}

\section{SUMMARY AND DISCUSSION}

We considered the four-dimensional accelerating BHs in AdS and explored the universal relation by analyzing
perturbative corrections to both the entropy and extremality bounds of the BHs.
It was found that the universal relation of accelerating BHs also holds
through an appropriate matching condition between the perturbation parameter $\epsilon$ and the parameter $K$.
After applying the matching condition, we obtained the parameter $K$($\epsilon$, $A^2l^2$) Eq.(\ref{K}) as the function of
the perturbation parameter $\epsilon$ and the variable $A^2l^2$.
Then, it was showed that $K$ grows up as $A^2 l^2$ increases with the given $\epsilon$ and
as $\epsilon$ increases with the fixed $A^2 l^2$ as illustrated in FIG. 1, and  FIG. 2.
In particular, when the acceleration horizons disappear,
$K\rightarrow\infty$ as $A^2l^2$ approaches 1 and string tension becomes infinity.
This implies that this BH experiences infinite acceleration due to this tension.
Thus, the universal relation seems not to work in the limit.
Furthermore, accelerating BHs with the perturbation parameter reduce pure AdS BH without conical deficits
since $K$ becomes $1$ when $A^2l^2=0$. Therefore, for the accelerating BHs in AdS,
the universal relation seems to hold within a constrained range $1\leq K<\infty$ ($0\leq A^2l^2<1$).

Recently, in the context of the holographic description, a study has been conducted on accelerating BHs
in three-dimensional AdS using an ADM decomposition to identify boundary data~\cite{Arenas-Henriquez:2023hur}.
It has been shown that a holographic conformal field theory resides
in a fixed curved background described by the holographic stress tensor of a perfect fluid.
It has been shown that a holographic conformal field theory resides in a fixed curved background described
by the holographic stress tensor of a perfect fluid. From the computation of entanglement entropy,
it has been found that increased acceleration reduces both the accessible region of the conformal boundary
and the entanglement entropy, indicating a loss of information in the dual theory.
It may be also intriguing to extend the study to the holographic accelerating BHs in three-dimensional AdS
in the context of the universal relation.

\section*{Acknowledgements}
This paper was supported by Wonkwang University in 2024.


\begin{thebibliography}{100}
\bibitem{Vafa:2005ui}
C.~Vafa,
[arXiv:hep-th/0509212 [hep-th]].

\bibitem{Arkani-Hamed:2006emk}
N.~Arkani-Hamed, L.~Motl, A.~Nicolis and C.~Vafa,
JHEP \textbf{06} (2007), 060
doi:10.1088/1126-6708/2007/06/060
[arXiv:hep-th/0601001 [hep-th]].

\bibitem{Banks:2010zn}
T.~Banks and N.~Seiberg,
Phys. Rev. D \textbf{83} (2011), 084019
doi:10.1103/PhysRevD.83.084019
[arXiv:1011.5120 [hep-th]].

\bibitem{Harlow:2022ich}
D.~Harlow, B.~Heidenreich, M.~Reece and T.~Rudelius,
Rev. Mod. Phys. \textbf{95} (2023) no.3, 3
doi:10.1103/RevModPhys.95.035003
[arXiv:2201.08380 [hep-th]].

\bibitem{Cheung:2018cwt}
C.~Cheung, J.~Liu and G.~N.~Remmen,
JHEP \textbf{10}, 004 (2018)
doi:10.1007/JHEP10(2018)004
[arXiv:1801.08546 [hep-th]].

\bibitem{Kats:2006xp}
Y.~Kats, L.~Motl and M.~Padi,
JHEP \textbf{12}, 068 (2007)
doi:10.1088/1126-6708/2007/12/068
[arXiv:hep-th/0606100 [hep-th]].

\bibitem{Reall:2019sah}
H.~S.~Reall and J.~E.~Santos,
JHEP \textbf{04}, 021 (2019)
doi:10.1007/JHEP04(2019)021
[arXiv:1901.11535 [hep-th]].

\bibitem{Cheung:2019cwi}
C.~Cheung, J.~Liu and G.~N.~Remmen,
Phys. Rev. D \textbf{100}, no.4, 046003 (2019)
doi:10.1103/PhysRevD.100.046003
[arXiv:1903.09156 [hep-th]].

\bibitem{Ma:2023qqj}
L.~Ma, Y.~Pang and H.~Lu,
JHEP \textbf{06}, 087 (2023)
doi:10.1007/JHEP06(2023)087
[arXiv:2304.08527 [hep-th]].



\bibitem{Goon:2019faz}
G.~Goon and R.~Penco,
Phys. Rev. Lett. \textbf{124}, no.10, 101103 (2020)
doi:10.1103/PhysRevLett.124.101103
[arXiv:1909.05254 [hep-th]].

\bibitem{Wei:2020bgk}
S.~W.~Wei, K.~Yang and Y.~X.~Liu,
Nucl. Phys. B \textbf{962}, 115279 (2021)
doi:10.1016/j.nuclphysb.2020.115279
[arXiv:2003.06785 [gr-qc]].

\bibitem{Chen:2020bvv}
D.~Chen, J.~Tao and P.~Wang,
Chin. Phys. C \textbf{45}, no.2, 025108 (2021)
doi:10.1088/1674-1137/abcf21
[arXiv:2004.10459 [gr-qc]].

\bibitem{Chen:2020rov}
Q.~Chen, W.~Hong and J.~Tao,
[arXiv:2005.00747 [gr-qc]].

\bibitem{Ma:2020xwi}
L.~Ma, Y.~Z.~Li and H.~Lu,
JHEP \textbf{01}, 201 (2021)
doi:10.1007/JHEP01(2021)201
[arXiv:2009.00015 [hep-th]].


\bibitem{Sadeghi:2020ciy}
J.~Sadeghi, S.~N.~Gashti, I.~Sakalli and B.~Pourhassan,
Nucl. Phys. B \textbf{1004}, 116581 (2024)
doi:10.1016/j.nuclphysb.2024.116581
[arXiv:2011.05109 [gr-qc]].

\bibitem{Sadeghi:2020xtc}
J.~Sadeghi, B.~Pourhassan, S.~Noori Gashti, S.~Upadhyay and E.~N.~Mezerji,
Phys. Scripta \textbf{98}, no.2, 025305 (2023)
doi:10.1088/1402-4896/acb40b
[arXiv:2011.14366 [physics.gen-ph]].

\bibitem{McInnes:2021frb}
B.~McInnes,
Nucl. Phys. B \textbf{971} (2021), 115525
doi:10.1016/j.nuclphysb.2021.115525
[arXiv:2104.07373 [gr-qc]].

\bibitem{McInnes:2021zlt}
B.~McInnes,
JHEP \textbf{12} (2021), 155
doi:10.1007/JHEP12(2021)155
[arXiv:2108.05686 [gr-qc]].

\bibitem{Noumi:2022ybv}
T.~Noumi and H.~Satake,
JHEP \textbf{12} (2022), 130
doi:10.1007/JHEP12(2022)130
[arXiv:2210.02894 [hep-th]].

\bibitem{Etheredge:2022rfl}
M.~Etheredge and B.~Heidenreich,
JHEP \textbf{12} (2023), 174
doi:10.1007/JHEP12(2023)174
[arXiv:2211.09823 [hep-th]].

\bibitem{Ko:2023nim}
J.~Ko and B.~Gwak,
JHEP \textbf{03} (2024), 072
doi:10.1007/JHEP03(2024)072
[arXiv:2312.17014 [gr-qc]].

\bibitem{Ko:2024ymy}
J.~Ko and B.~Gwak,
[arXiv:2406.10567 [gr-qc]].



\bibitem{Cano:2019ycn}
P.~A.~Cano, S.~Chimento, R.~Linares, T.~Ort\'\i{}n and P.~F.~Ram\'\i{}rez,
JHEP \textbf{02} (2020), 031
doi:10.1007/JHEP02(2020)031
[arXiv:1910.14324 [hep-th]].

\bibitem{Cremonini:2019wdk}
S.~Cremonini, C.~R.~T.~Jones, J.~T.~Liu and B.~McPeak,
JHEP \textbf{09} (2020), 003
doi:10.1007/JHEP09(2020)003
[arXiv:1912.11161 [hep-th]].

\bibitem{Arkani-Hamed:2021ajd}
N.~Arkani-Hamed, Y.~t.~Huang, J.~Y.~Liu and G.~N.~Remmen,
JHEP \textbf{03} (2022), 083
doi:10.1007/JHEP03(2022)083
[arXiv:2109.13937 [hep-th]].

\bibitem{Sadeghi:2020ntn}
J.~Sadeghi, S.~Noori Gashti and E.~Naghd Mezerji,
Phys. Dark Univ. \textbf{30}, 100626 (2020)
doi:10.1016/j.dark.2020.100626

\bibitem{McPeak:2021tvu}
B.~McPeak,
Phys. Rev. D \textbf{105} (2022) no.8, L081901
doi:10.1103/PhysRevD.105.L081901
[arXiv:2112.13433 [hep-th]].

\bibitem{Sadeghi:2022xcr}
J.~Sadeghi, B.~Pourhassan, S.~Noori Gashti and S.~Upadhyay,
Annals Phys. \textbf{447}, 169168 (2022)
doi:10.1016/j.aop.2022.169168
[arXiv:2201.04071 [gr-qc]].

\bibitem{Xiao:2023two}
Y.~Xiao and Y.~Y.~Liu,
[arXiv:2312.07127 [gr-qc]].



\bibitem{Kinnersley:1970zw}
W.~Kinnersley and M.~Walker,
Phys. Rev. D \textbf{2}, 1359-1370 (1970)
doi:10.1103/PhysRevD.2.1359

\bibitem{Plebanski:1976gy}
J.~F.~Plebanski and M.~Demianski,
Annals Phys. \textbf{98}, 98-127 (1976)
doi:10.1016/0003-4916(76)90240-2

\bibitem{Griffiths:2005qp}
J.~B.~Griffiths and J.~Podolsky,
Int. J. Mod. Phys. D \textbf{15}, 335-370 (2006)
doi:10.1142/S0218271806007742
[arXiv:gr-qc/0511091 [gr-qc]].

\bibitem{Anabalon:2018ydc}
A.~Anabal\'on, M.~Appels, R.~Gregory, D.~Kubiz\v{n}\'ak, R.~B.~Mann and A.~Ovg\"un,
Phys. Rev. D \textbf{98}, no.10, 104038 (2018)
doi:10.1103/PhysRevD.98.104038
[arXiv:1805.02687 [hep-th]].

\bibitem{Belhaj:2020lai}
A.~Belhaj, H.~El Moumni and K.~Masmar,
Adv. High Energy Phys. \textbf{2020} (2020), 4092730
doi:10.1155/2020/4092730
[arXiv:2008.00931 [hep-th]].

\bibitem{Zhang:2018hms}
J.~Zhang, Y.~Li and H.~Yu,
JHEP \textbf{02} (2019), 144
doi:10.1007/JHEP02(2019)144
[arXiv:1808.10299 [hep-th]].

\bibitem{Arenas-Henriquez:2023hur}
G.~Arenas-Henriquez, A.~Cisterna, F.~Diaz and R.~Gregory,
JHEP \textbf{09} (2023), 122
doi:10.1007/JHEP09(2023)122
[arXiv:2308.006133 [hep-th]].

\bibitem{Cisterna:2023qhh}
A.~Cisterna, F.~Diaz, R.~B.~Mann and J.~Oliva,
JHEP \textbf{11} (2023), 073
doi:10.1007/JHEP11(2023)073
[arXiv:2309.05559 [hep-th]].

\bibitem{Tian:2023ine}
J.~Tian and T.~Lai,
[arXiv:2312.13718 [hep-th]].



\bibitem{Podolsky:2002nk}
J.~Podolsky,
Czech. J. Phys. \textbf{52}, 1-10 (2002)
doi:10.1023/A:1013961411430
[arXiv:gr-qc/0202033 [gr-qc]].

\bibitem{Hong:2003gx}
K.~Hong and E.~Teo,
Class. Quant. Grav. \textbf{20}, 3269-3277 (2003)
doi:10.1088/0264-9381/20/14/321
[arXiv:gr-qc/0305089 [gr-qc]].















\bibitem{Appels:2017xoe}
M.~Appels, R.~Gregory and D.~Kubiznak,
JHEP \textbf{05}, 116 (2017)
doi:10.1007/JHEP05(2017)116
[arXiv:1702.00490 [hep-th]].






\end{thebibliography}
\end{document}